\documentclass[a4paper,twocolumn,11pt]{article}
\usepackage{graphics}
\usepackage{amsmath,amssymb}
\usepackage{color}

\newcommand{\DI}{\ensuremath{D_\mathrm{I}}}
\newcommand{\chiSG}{\ensuremath{\chi_\mathrm{SG}}}
\newcommand{\TSG}{\ensuremath{T_\mathrm{f}}}
\newcommand{\Tc}{\ensuremath{T_\mathrm{c}}}

\newcommand{\Ns}{\ensuremath{N_\mathrm{s}}}
\newcommand{\qEA}{\ensuremath{q_\mathrm{EA}^2}}
\newcommand{\valpha}{\ensuremath{\vec{\alpha}}}
\newcommand{\vQ}{\ensuremath{\vec{Q}}}

\begin{document}
%\title{\bf Numerical studies on spin-glass transition in bond-disordered pyrochlore Heisenberg antiferromagnets by the extended loop algorithm} 
\title{\bf Extended loop algorithm for pyrochlore Heisenberg spin models with spin-ice type degeneracy: application to spin-glass transition in antiferromagnets coupled to local lattice distortions}
\author{
     Hiroshi SHINAOKA \\ 
{\it Nanosystem Research Institute (NRI), }\\
{\it National Institute of Advanced Industrial Science and Technology (AIST),}\\
{\it 1-1-1 Umezono, Tsukuba, Ibaraki 305-8568}\\
}
\date{}
\maketitle\thispagestyle{empty}

\section*{Abstract}
\vspace{-0.6em}
For Ising spin models which bear the spin-ice type macroscopic (quasi-)degeneracy, conventional classical Monte Carlo (MC) simulation using single spin flips suffers from dynamical freezing at low temperatures ($T$).
A similar difficulty is seen also in a family of Heisenberg spin models with easy-axis anisotropy or biquadratic interactions.
In the Ising case, the difficulty is avoided by introducing a non-local update based on the loop algorithm.
We present an extension of the loop algorithm to the Heisenberg case.
As an example of its application, we review our recent study on spin-glass (SG) transition in a bond-disordered Heisenberg antiferromagnet coupled to local lattice distortions.

\section{Introduction}
Recently, increasing attention has been devoted to low-temperature behavior of geometrically frustrated magnets~\cite{Diep05}.
Spin glass, in which spins are frozen randomly, is one of low-$T$ phases widely observed in geometrically frustrated materials.
However, it is unclear so far how the nature of SG is different from the canonical one driven solely by randomness.

\begin{figure}[ht]
 \centering
 \resizebox{0.5\textwidth}{!}{\includegraphics{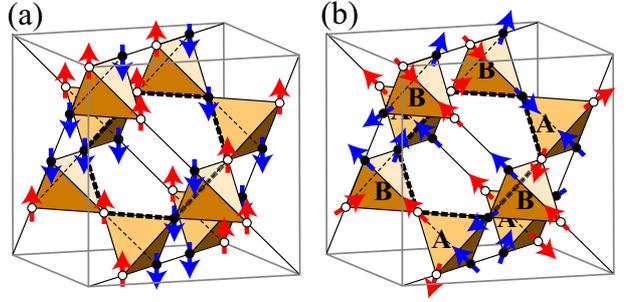}}
 \caption{
 A 16-site cubic unit cell of the pyrochlore lattice is shown with Ising spins along a global axis [(a)] and local $\langle 111 \rangle$ axes [(b)].
Spins are denoted by arrows.
(a) White circles represent spins pointing upward, while black circles the opposite.
The spin configuration is an example of the spin-ice type states. % in which the `two-up two-down' local constraint is satisfied in every tetrahedron.
The hexagon with a bold dashed line denotes an example of loops with alternating black and white sites.
(b) The ice-rule configuration equivalent to (a) is shown with Ising spins along the local $\langle 111 \rangle$ axes. A and B represent two different types of tetrahedra.
Black circles represent spins pointing inward in terms of type-A tetrahedra, while white circles the opposite.
}
 \label{fig:pyrochlore}
\end{figure}

An antiferromagnet on a pyrochlore lattice (Fig.~\ref{fig:pyrochlore}) is a typical example of geometrically frustrated spin systems.
Recently, several puzzling SG behaviors have been pointed out experimentally in pyrochlore-based magnets.
One of the surprising aspects is that, in these SG materials, the SG transition temperature $\TSG$ appears to be almost independent of the strength of disorder $\Delta$; e.g., for (La$_x$Y$_{1-x}$)$_2$Mo$_2$O$_7$, $\TSG\simeq 22\text{K}$ stays almost constant for $x\le 0.5$~\cite{Greedan86}.
Similar plateau behavior of $\TSG$ is also observed in (Zn$_{1-x}$Cd$_x$)Cd$_2$O$_4$~\cite{Ratcliff02}.
Another distinctive aspect is that, for these SG materials, $\TSG$ is much higher than a theoretically expected value for a moderate strength of disorder $\Delta$~\cite{Saunders07, Andreanov10,Tam10}.
These behaviors suggest that some important factor is missing in the previous SG theories: A candidate is the magnetoelastic coupling.
For example, various microscopic probes have pointed out importance of local lattice distortions in Y$_2$Mo$_2$O$_7$ although this material exhibits no uniform lattice distortion.
They are crucial also in (Zn$_{1-x}$Cd$_x$)Cr$_2$O$_4$ because it exhibits the spin-lattice ordering at $x=0$.

Motivated by these puzzles, we recently investigated effects of the magnetoelastic coupling on the spin-glass transition by considering the following classical Heisenberg spin model~\cite{Shinaoka10b}:
\begin{equation}
	\mathcal{H}= \sum_{\langle i,j \rangle} 
	\Big[ J_{ij} \vec{S}_i \cdot \vec{S}_j - b_{ij} 
	\big( \vec{S}_i \cdot \vec{S}_j 
	\big)^2 
	\Big], \label{eq:ham-b2}
\end{equation}
where $\vec{S}_i$ denotes a Heisenberg spin at site $i$ and the sum runs over nearest-neighbor bonds.
The biquadratic interaction $b_{ij}~(\equiv bJ_{ij},~b>0)$ is induced by the spin-lattice coupling to local lattice distortions~\cite{Penc04}.
Note that such `ferro'-type biquadratic interaction favors collinear spin configurations. 
We introduce static bond disorder as a uniformly-distributed randomness as $J_{ij} \in [J-\Delta,J+\Delta]$ with $0\le \Delta < J$. 
At $\Delta=0$, this model exhibits a nematic transition at $\Tc \sim b$, below which spins select a common axis without selecting their directions on it; the system remains magnetically disordered down to zero $T$.
The ground-state degeneracy is equivalent to that of a nearest-neighbor Ising antiferromagnet on a pyrochlore lattice [see Fig.~\ref{fig:pyrochlore}(a)].
The ground-state degenerate manifold of this Ising model is identified by a set of local constraints enforcing two spins pointing up and two spins pointing down in every tetrahedron~\cite{Anderson56}.
This is called the ice rule because of an analogy to the constraint on positions of protons in hexagonal ice~\cite{Bernal33,Pauling35}.
Similar situation is seen also in the so-called spin-ice model in which Ising spins along the local $\langle 111 \rangle$ axes interact with each other ferromagnetically~\cite{Harris97,Ramirez99} [see Fig.~\ref{fig:pyrochlore}(b)]. 
In these systems, the degenerate configurations are separated by energy barriers of the order of the dominant interaction scale $J$, and the standard single-spin-flip algorithm does not work at low $T \ll J$ in classical MC simulation.
%The difficulty remains even when the Ising discreteness is relaxed and spins can fluctuate, as long as the ground-state manifold retains a multivalley structure.
The difficulty remains even when the Ising discreteness is relaxed and spins can fluctuate as in the present case of the biquadratic interaction, as long as the ground-state manifold retains a multivalley structure.
Indeed, the single-spin-flip algorithm does not work at low $T\ll b$ for the model (\ref{eq:ham-b2}) even at $\Delta=0$.
A similar situation is seen also in classical Heisenberg models with single-ion easy-axis anisotropy.

In the Ising case, the difficulty is avoided by introducing a global flip called the loop flip, in which one reverses all Ising spins on a specific closed loop passing through tetrahedra~\cite{Melko04}; the loop is chosen so that the spins are up and down (or inward and outward) alternatively along the loop as illustrated in Fig.~\ref{fig:pyrochlore}.
The loop flip connect different ice-rule states bypassing the energy barriers.
In the Heisenberg case, however, it is nontrivial how to define the loop with alternating spins.
Moreover, the loop flip procedure is not unique because of the continuous degrees of freedom.
These argued us to extend the loop algorithm to the Heisenberg case to investigate low-$T$ properties of the model~(\ref{eq:ham-b2}).

This report is organized as follows.
In \S~\ref{sec:lm-ising}, we briefly review the loop algorithm for Ising models~\cite{Melko04}.
Section~\ref{sec:elm} is devoted to the extension of the loop algorithm to Heisenberg spin systems and its benchmarks~\cite{Shinaoka10a,Shinaoka11a}.
In \S~\ref{sec:appl}, we report results of our numerical study on the model~(\ref{eq:ham-b2}) using the extended algorithm.

\vspace{-1.5em}
\section{Loop algorithm for Ising spin systems}\label{sec:lm-ising}
Before considering an extension of the loop algorithm to Heisenberg spin systems, here we briefly review the loop algorithm for Ising spin models~\cite{Melko04}. To generalize the following discussion, we assign black and white to two degrees of freedom of Ising spins in an appropriate manner.
For example, black and white simply correspond to up and down spins, respectively, for the antiferromagnetic Ising model [Fig.~\ref{fig:pyrochlore}(a)].
For the spin ice model~[Fig.~\ref{fig:pyrochlore}(b)], black and white represent inward and outward spins in terms of type-A tetrahedra, respectively.
Then the loop flip consists of two steps; first, we identify a closed loop which consists of alternating alignment of black and white sites, and then we try to flip all Ising spins on the loop. 

\begin{figure}[ht]
 \centering
 \resizebox{0.35\textwidth}{!}{\includegraphics{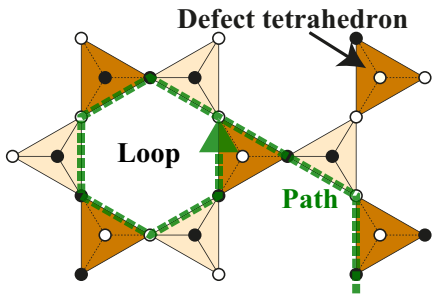}}
 \caption{Schematic picture for a loop construction by tracing a path through ice-rule tetrahedra. The path is made of alternating black and white sites. For simplicity, the figure shows a $\langle 111 \rangle$ kagome layer with connected tetrahedra. The path is denoted by a dashed line, and its left part represents an example of a closed loop. }
 \label{fig:short-loop}
\end{figure}

Such a closed loop can be constructed by using the short loop algorithm~\cite{Melko04}.
In the short loop algorithm, one traces a path through alternating black and white sites in ice-rule tetrahedra.
A loop is formed when the path encounters any tetrahedron already included in the path as illustrated in Fig.~\ref{fig:short-loop}.
At finite $T$, thermal fluctuations induce ``defect tetrahedra" in which the ice-rule condition is violated.
To maintain detailed balance, the path must be traced so that it does not involve defect tetrahedra on it.

After the construction of a closed loop, all colors on the loop are reversed simultaneously by flipping the spins.
When all the ice-rule states are energetically degenerate, the flip is always accepted (rejection free) in the MC sampling because the flip does not the total energy.
When there are residual interactions which lift the degeneracy, the loop flip is accepted according to the Metropolis criterion.

At finite $T$, the loop flip update does not satisfy ergodicity because it changes neither the spin configurations in defect tetrahedra nor the number of defect tetrahedra. It is, therefore, necessary to use the loop flip together with another update such as the standard single-spin flip for retaining the ergodicity.
%This is easily achieved by introducing the standard single-spin flip in MC samplings.

\section{Extended Loop algorithm}\label{sec:elm}
Now we extend the loop algorithm to Heisenberg spin systems with the spin-ice type degeneracy: 
Heisenberg spin systems with (1) single-ion anisotropy~\cite{Shinaoka10a} and (2) biquadratic interactions~\cite{Shinaoka11a}.
Section~\ref{sec:overview-extension} is devoted to an overview of the extended loop algorithm.
In \S~\ref{sec:single-ion} and \S~\ref{sec:biquadratic}, we review the detailed procedure of the algorithm for the cases (1) and (2), respectively.
Benchmark results are also given.

\subsection{Overview of the extended algorithm}\label{sec:overview-extension}
The extended loop algorithm consists of the following three steps:
\begin{enumerate}
	\item We first project the Heisenberg spin $\vec{S}_i$ onto an appropriate projection axis $\valpha_i$ to assign black and white colors at every site $i$.
 \item Then we construct a loop consisting of alternating black and white sites. 
 \item All colors on the constructed loop are reversed simultaneously.
\end{enumerate}

In the case of the single-ion anisotropy, the projection axis $\valpha_i$ is simply given by the easy axis at site $i$.
For example, in the case of antiferromagnets with easy-axis anisotropy along the $z$ axis, which are extensions of the Ising antiferromagnet~[Fig.~\ref{fig:pyrochlore}(a)], we set $\vec{\alpha}_i = (0,0,1)$ for all the sites.
While, in the case of ferromagnets with the local $\langle 111 \rangle$ anisotropy, which are natural extensions of the spin ice model~[Fig.~\ref{fig:pyrochlore}(b)], we set $\vec{\alpha}_i$ to the direction connecting the centers of neighboring tetrahedra from type B to A.
In the case of the biquadratic interactions, however, the systems retain $O$(3) spin rotational symmetry and have no explicit anisotropy axis to project the spins on. 
Therefore, it is necessary to deduce the common axis selected by spins for each MC sample.
In \S~\ref{sec:biquadratic}, we introduce a simple way to determine the projection axis.

Once the projection axis $\valpha_i$ is defined at every site, we assign black and white colors to sites at which $\vec{S}_i \cdot \vec{\alpha}_i \ge 0$ and $\vec{S}_i \cdot \vec{\alpha}_i < 0$, respectively.
Based on this definition, we can construct a closed loop with alternating black and white sites by following the short loop algorithm similarly to the Ising case.

\begin{figure}[ht]
 \centering
 \resizebox{0.4\textwidth}{!}{\includegraphics{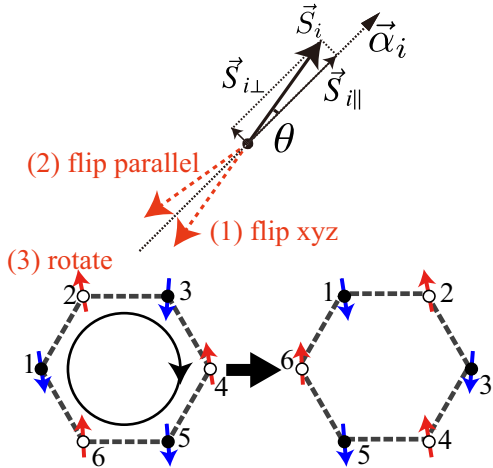}}
 \caption{Different ways to reverse black and white: (1) \textit{flip xyz}, (2) \textit{flip parallel}, and (3) \textit{rotate}. $\vec{\alpha}_i$ is the projection axis at site $i$. The \textit{rotate} is applicable to systems with global anisotropy axes. See the text for details.}
 \label{fig:loop-flip}
\end{figure}

In the step 3, as mentioned above, the way to reverse black and white is not unique because of the continuous degrees of freedom.
Three different ways are illustrated in Fig.~\ref{fig:loop-flip}: (1) \textit{flip xyz}, (2) \textit{flip parallel} and (3) \textit{rotate}.
In \textit{flip xyz}, all three Cartesian components of $\vec{S}_i$ are reversed as $\vec{S}_i \rightarrow - \vec{S}_i$, while in \textit{flip parallel}, only components parallel to the easy axes, $\vec{S}_{i\parallel}$, are reversed as $\vec{S}_i \rightarrow \vec{S}_i  - 2 (\vec{S}_i \cdot \valpha_i) \valpha_i$.
In \textit{rotate}, which is applicable to systems with global anisotropy axes, one translates every spin to the neighboring site on the loop simultaneously in the same direction. %in which the loop was formed.

For models which retain the ground-state spin-ice type degeneracy, one might expect that these updates become equivalent and always accepted at low $T$;
this is naively expected since thermal fluctuations vanish and all the ice-rule configurations with spins parallel to the easy axes become energetically degenerate.
However, this is not the case:
%Even at low $T$, spins thermally fluctuate around the easy axes\blue{, contributing to the total energy change in the loop flips in different ways depending on the model.}
As discussed in the following sections, careful consideration on the energy change is necessary to choose an efficient method.

\begin{figure}[ht]
 \centering
 \resizebox{0.475\textwidth}{!}{\includegraphics{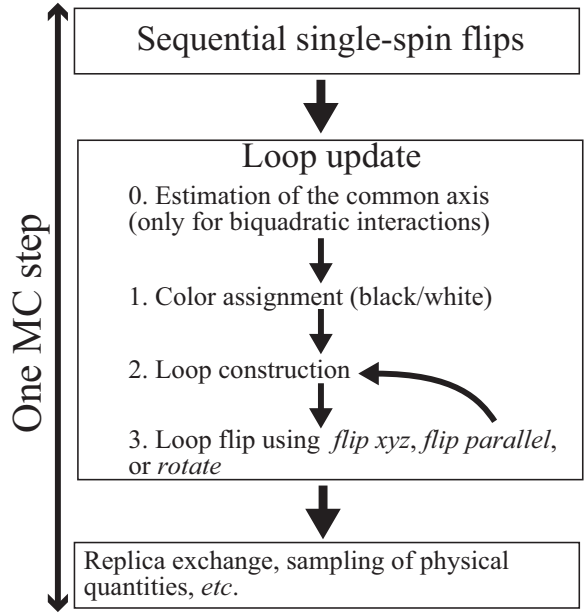}}
 \caption{Flowchart of our computational code with the extended loop update, the single-spin update, and the replica exchange MC method.}
 \label{fig:flowchart}
\end{figure}

Now we comment on technical aspects of implementation of the algorithm.
We implemented a computation code based on the extended loop algorithm, single-spin-flip algorithm and exchange MC method~\cite{Hukushima96}, which is used in the following MC simulations.
As illustrated in Fig.~\ref{fig:flowchart}, one MC step consists of a sweep of the lattice by sequential single-spin flips, followed by the loop update and replica exchange between neighboring temperatures.
%In the single-spin flips, we randomly choose a new spin state on the unit sphere.
The loop flips are repeated until the number of tetrahedra visited in the loop construction exceeds the number of lattice sites.
The loop-flip section takes CPU time comparable to the single-spin-flip sweep.
%a sweep of the lattice sites by single-spin flips.

\subsection{Heisenberg spin systems with single-ion anisotropy}\label{sec:single-ion}
In this section, we review the loop algorithm extended to Heisenberg models with easy-axis anisotropy~\cite{Shinaoka10a}.
We start with a simple Hamiltonian:
\begin{equation}
	\mathcal{H}= J \sum_{\langle i,j\rangle} \vec{S}_i \cdot \vec{S}_j -
 \DI \sum_i \left(\vec{S}_i \cdot \vec{\alpha}_i\right)^2, \label{eq:ham-DI}
\end{equation}
where $\vec{S}_i$ denotes a classical Heisenberg spin at site $i$ (we take $|\vec{S}_i|=1$) and $\DI$ $(>0)$ is the single-ion easy-axis anisotropy.
The easy axis $\vec{\alpha}_i~(|\valpha_i|=1)$ defines the projection axis in the loop update.
Although the exchange interaction is limited to nearest neighbors for simplicity here, the following algorithm is applicable to more general models with farther-neighbor or bond-dependent interactions.
We consider periodic systems of cubic geometry with $L^3$ unit cells with totally $\Ns = 16L^3$ spins.
We take the energy unit as $|J|=1$.

For models which retain the spin-ice type degeneracy in the ground state, spins fluctuate around the easy axes by angles of $O(\sqrt{T})$ at low $T\ll \DI$.
Under the influence of the thermal fluctuations, the energy change in the loop flip is estimated as follows\footnote{For more detailed discussion, refer to ref.~\cite{Shinaoka10a} and a short note in reference 22 of Ref.~\cite{Shinaoka11a}.}:
\begin{align}
	\Delta E &\propto T &\text{(\textit{flip xyz})},\nonumber\\
	\Delta E &\propto T^2 &\text{(\textit{flip parallel})}. \nonumber
\end{align}
Because the acceptance rate of the Metropolis algorithm is given by $\mathrm{min} \{1, \exp(-\Delta E/T)\}$, thermal fluctuations are irrelevant for \textit{flip parallel} in the sense that $\lim_{T\rightarrow 0} \exp(- \Delta E/T) = 1$. On the contrary, thermal fluctuations are relevant for \textit{flip xyz} since $\lim_{T\rightarrow 0} \exp(- \Delta E/T)  < 1$.
This consideration indicates that \textit{flip parallel} becomes refection free in the limit of $T\rightarrow 0$, while \textit{flip xyz} not even for models which retain the spin-ice type degeneracy.

\begin{figure}[ht]
 \centering
 \resizebox{0.4\textwidth}{!}{\includegraphics{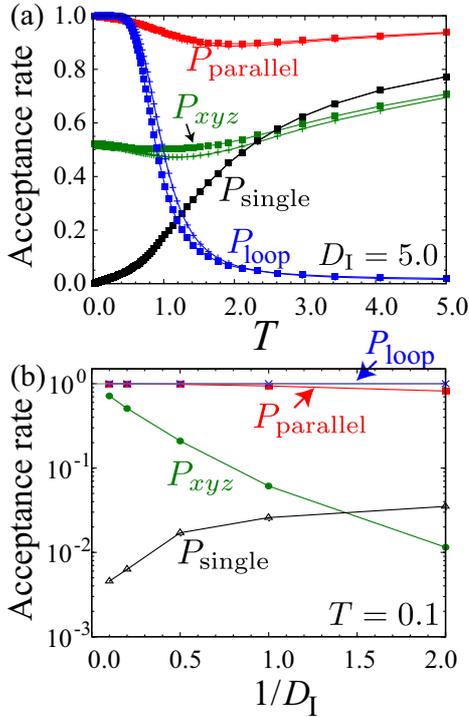}}
 \caption{(a) Temperature dependences of the acceptance rates of the single-spin flip ($P_{\text{single}}$), the probability of formation of closed loops ($P_{\text{loop}}$), the acceptance rates of flip of a formed loop by \textit{flip xyz} ($P_{xyz}$) and by \textit{flip parallel} ($P_{\text{parallel}}$). The data are calculated for the model (\ref{eq:ham-DI}) at $\DI=5.0$ with the antiferromagnetic exchange interaction $J=1$ and $\valpha_i=(0,0,1)$. The data for $L=2$ and $L=4$ are denoted by crosses and filled squares, respectively. (b) $\DI$ dependence of $P_{\text{parallel}}$, $P_{xyz}$, $P_{\text{loop}}$, and $P_{\text{single}}$ at $T=0.1$ for $L=2$.}
 \label{fig:demo-D5.0}
\end{figure}

The efficiency of the loop flip is demonstrated in Fig.~\ref{fig:demo-D5.0}(a) for the model~(\ref{eq:ham-DI}) with the antiferromagnetic exchange interaction $J=1$ and $\valpha_i=(0,0,1)$.
This model retains the spin-ice type degeneracy in the ground state.
At low $T\ll|J|=1$, spin configurations are gradually enforced to satisfy the ice rule, and the acceptance rate of the single-spin flip, $P_{\text{single}}$, is suppressed below $T \sim |J|$ and vanishes in the low-$T$ limit. On the contrary, the acceptance rate of loop flips increases at low $T$. As shown in Fig.~\ref{fig:demo-D5.0}(a), the probability that a closed loop is successfully formed, $P_\mathrm{loop}$, steeply increases below $T \sim |J|$, indicating that almost all tetrahedra start to follow the ice rule below this temperature. At the same time, the acceptance rate of flips of a formed loop gradually increases at low $T < |J|$ and remains finite; here, $P_{xyz}$ and $P_{\text{parallel}}$ are the rate for \textit{flip xyz} and \textit{flip parallel}, respectively. The total acceptance rate of the loop flip is given by the product as $P_\mathrm{loop} \times P_{xyz}$ or $P_\mathrm{loop} \times P_{\text{parallel}}$, and it sharply increases at $T < |J|$, compensating the decrease of $P_{\text{single}}$. 

As clearly indicated in Fig.~\ref{fig:demo-D5.0}(a), the acceptance rate of \textit{flip parallel} is always larger than that of \textit{flip xyz}, being consistent with the above argument. In particular, $P_{\text{parallel}}$ approaches 1 (rejection free) as $T \to 0$, whereas $P_{xyz}$ goes to a smaller value $\sim 0.5$. The reduction of $P_{xyz}$ becomes larger for smaller anisotropy $\DI$. This is demonstrated at $T=0.1$ in Fig.~\ref{fig:demo-D5.0}(b); $P_{xyz}$ decreases almost exponentially with $1/\DI$.
On the other hand, $P_{\text{parallel}}$ is almost independent of $\DI$ and remains rejection free at $T\rightarrow 0$ in the wide range of $\DI$.

\subsection{Heisenberg spin systems with biquadratic interaction}\label{sec:biquadratic}
In this section, we review the loop algorithm extended to classical antiferromagnetic Heisenberg models with biquadratic interactions.
We start with a Hamiltonian of a simple form:
\begin{equation}
\mathcal{H}=\sum_{\langle i,j\rangle} \left\{J\left(\vec{S}_i \cdot \vec{S}_j\right) - b\left(\vec{S}_i \cdot \vec{S}_{j}\right)^2 \right\}, \label{eq:ham-b}
\end{equation}
where $b~(>0)$ is the biquadratic interaction.
The model (\ref{eq:ham-b2}) reduces to this model when $\Delta=0$.
We note that such `ferro'-type biquadratic interaction originates in quantum and thermal fluctuations as well as the spin-lattice coupling.
We consider the antiferromagnetic exchange interaction $J>0$, and take the energy unit as $J=1$.
The sum runs over nearest-neighbor bonds.
The following algorithm is applicable to more general models with farther-neighbor or bond-dependent interactions such as the model (\ref{eq:ham-b2}) with $\Delta>0$.

As mentioned above, it is necessary to deduce the common axis $\vQ$ selected by $b$ for each MC sample.
Here, we explain a simple way to determine the projection axis.
We first pick up a set of $N_\mathrm{T}$ tetrahedra \{$\mathcal{T}_m$\} ($m=1,\cdots,N_\mathrm{T}$) randomly from the whole system.
Starting from an initial guess $\valpha_0$ [we take $\valpha_0 = (0,0,1)$], the normalized projection axis $\valpha$ is obtained iteratively by
\begin{eqnarray}
	\valpha_{n+1} \propto \sum_{i \in \{ \mathcal{T}_m \}} \mathrm{sign} (\vec{S}_{i} \cdot \valpha_n) \vec{S}_{i}.\nonumber
\end{eqnarray}
Here the sum is taken over all spins belonging to the selected tetrahedra \{$\mathcal{T}_m$\}, and $n~(=0,1,\cdots,n_\mathrm{max}-1)$ is the index of the iteration.
For larger $N_\mathrm{T}$ and $n_\mathrm{max}$, the resultant $\valpha=\valpha_{n_\mathrm{max}}$ gives a better approximation of $\vQ$. In practice, we take $N_\mathrm{T} = 16$ and $n_\mathrm{max}=6$ for the system sizes $L\geq 3$ in the following MC simulations.
It should be noted that, to ensure the detailed balance, loops must be constructed avoiding the tetrahedra included in \{$\mathcal{T}_m$\} as well as defect tetrahedra in which the ice rule is violated: Otherwise, the loop flip becomes irreversible because the flip changes $\valpha$.
Because loop flips do not change $\valpha$, $\valpha$ is determined once at the beginning of each MC step as shown in Fig.~\ref{fig:flowchart}.
The computational cost for estimating $\valpha$ is negligible in practical calculations.

\begin{figure}[ht]
 \centering
 \resizebox{0.4\textwidth}{!}{\includegraphics{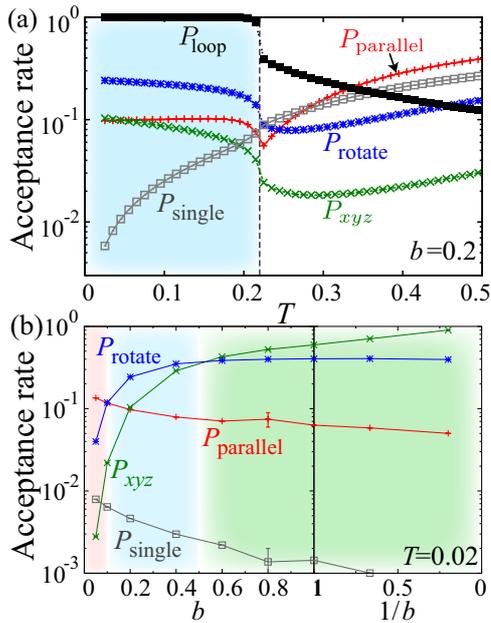}}
 \caption{(a) Temperature dependences of the acceptance rates. For the definition of the acceptance rates, see the text. The data are calculated at $b=0.2$ and $L=8$. The vertical broken line denotes the nematic transition temperature $\Tc$. (b) $b$ dependence of the acceptance rates at $T=0.02$. The most efficient method depends on the value of $b$.}
 \label{fig:b-benchmark}
\end{figure}

Now we show benchmark results in Fig~\ref{fig:b-benchmark}.
The number of spins in the system $\Ns$ is given by $16L^3$.
For $b>0$, the model~(\ref{eq:ham-b}) exhibits a nematic transition at $\Tc \sim b$, below which spins select a common axis.
At low $T$ compared to $b$ and $J$, spin configurations are enforced to satisfy the `two-up two-down' ice rule, and the acceptance rate of the single-spin flip, $P_{\mathrm{single}}$, is suppressed.
This is demonstrated in Fig.~\ref{fig:b-benchmark}(a) with $b=0.2$.
While, the probability that a closed loop is successfully formed, $P_\mathrm{loop}$, steeply increases below $\Tc \sim b$, indicating that almost all tetrahedra start to follow the ice rule below $\Tc$. 
The acceptance rate of flips of a formed loop also increases below $\Tc$ and remains finite as $T \to 0$; here, $P_{xyz}$, $P_{\mathrm{parallel}}$, and $P_\mathrm{rotate}$ are the rates for \textit{flip xyz}, \textit{flip parallel}, and \textit{rotate}, respectively.
The acceptance rate of the loop flip sharply increases at $T < \Tc$, compensating the decrease of $P_{\mathrm{single}}$. 

In contrast to the case of the single-ion anisotropy, the most efficient loop flip depends on the value of $b$ as demonstrated in Fig.~\ref{fig:b-benchmark}(b): $P_\mathrm{parallel}$ becomes most efficient as $b\to 0$, while, in the opposite limit (i.e., $b\to +\infty$), \textit{flip xyz} becomes rejection free but the other two not.
In the intermediate regime, i.e, $0.1<b<0.5$, \textit{rotate} is most efficient.
The difference of the efficiency of the loop flips are understood by the following consideration.
Considering a given state at a finite $T$ well below $\Tc$, its energy measured from the ground-state energy is given by $E=E_J + E_b=O(T)$, where $E_J$ and $E_b$ are the energies corresponding to the first and second terms in eq.~(\ref{eq:ham-b}), respectively. 
%Both $E_J$ and $E_b$ are of the order of $T$ at low $T$.
The three loop flips change the two contributions in different ways. 
The \textit{flip xyz} conserves $E_b$, while the other two not.
This is why the \textit{flip xyz} becomes most efficient at $b\to \infty$ where $E\simeq E_b$ and $|E_b|\gg |E_J|$.
In the opposite limit, the contribution of $E_J$ becomes dominating in the energy change because $|E_J|\gg |E_b|$.
The \textit{flip xyz}, \textit{flip parallel} and \textit{rotate} change $E_J$ by $O(T)$, $O(T^2)$ and $O(T)$, respectively.
This is why \textit{flip parallel} is most efficient at small $b$.
The complete argument is given in ref.~\cite{Shinaoka11a}.

\section{Application}\label{sec:appl}
Now, we review results of our MC study on the model (\ref{eq:ham-b2}) utilizing the extended loop algorithm and discuss the obtained $\Delta$-$T$ phase diagram.
When $b=0$, the model reduces to the previously studied one~\cite{Saunders07, Andreanov10,Tam10, Bellier-Castella01}.
In this case, by turning on $\Delta$, the SG transition appears roughly linearly as $\TSG \simeq 0.1\Delta$~\cite{Saunders07, Andreanov10,Tam10}.
As already mentioned, for a finite spin-lattice coupling $b>0$, the present model exhibits a nematic transition at $\Tc \sim b$ for $\Delta=0$.
Our interest is how the SG transition appears in the presence of $b$ by turning on $\Delta$.

In the following MC simulations, we take $b=0.2$ and employ \textit{rotate}, which is the most efficient flip at $b=0.2$.
We also adopt the exchange MC method~\cite{Hukushima96} and the overrelaxation update~\cite{Alonso96} to further accelerate MC dynamics.
To identify the SG transition, we calculate the SG susceptibility $\chiSG\equiv \Ns \qEA$, where the Edwards-Anderson order parameter $\qEA$~\cite{Edwards75} is defined as the overlap of two independent replicas with the same interaction set $\{J_{ij}\}$.

The computational code is parallelized by using the message passing interface (MPI) library.
The MPI processes are divided into independent groups which handle different $\{J_{ij}\}$.
In the replica exchange, spin configurations are swapped within a process or across neighboring processes.
In the following simulation, we take four independent replica for each interaction set.
For $L=5$, a MC run of $1.2\times 10^7$ steps with one interaction set and 64 temperatures takes 10 fours using 64 CPU cores in System B of the Supercomputer Center (ISSP, Univ. of Tokyo).
\begin{figure}[ht]
 \centering
 \resizebox{0.425\textwidth}{!}{\includegraphics{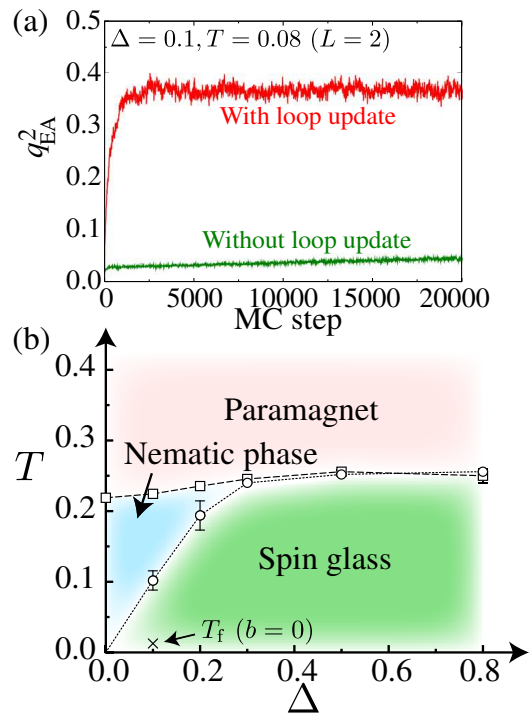}}
 \caption{(a) Comparison of thermalization processes of $\qEA$ with/without the loop update (\textit{rotate}). (b) Calculated $\Delta$-$T$ phase diagram of the model~(\ref{eq:ham-b2}).}
 \label{fig:pd}
\end{figure}

In Fig.~\ref{fig:pd}(a), we compare thermalization processes of $\qEA$ with/without the loop update at $\Delta=0.1$ and $T=0.08$ for the system size $L=2$ with 128 spins.
For the both cases, we took 16 temperatures in the range of $0.08\le T\le 0.2$ for parallel tempering.
The MC dynamics without the loop update exhibits a severe freezing, and does not reach thermal equilibrium after $2\times 10^4$ MC steps.
However, the thermalization process is greatly accelerated by the loop update and $\qEA$ converges to the thermal-equilibrium value after $2.5\times 10^3$ MC steps.
These clearly show the advantage of the loop algorithm in investigating the low-$T$ properties of the present model.
The advantage becomes more pronounced for larger $L$, because the number of the ice-rule states grows as $L$ increases.

Figure~\ref{fig:pd}(b) presents the phase diagram obtained by MC simulation with the loop update for $L\le 5$.
By introducing the disorder $\Delta$, the SG transition appears at a finite $\TSG$.
In the weakly-disordered region, i.e., $\Delta \lesssim b$, $\TSG$ is roughly proportional to $\Delta$ as $\TSG\simeq \Delta$.
A remarkable point is that $\TSG$ is largely enhanced by $b$~\cite{Saunders07, Andreanov10, Tam10}:
The enhancement factor is, e.g., about 5-10.
At $\Delta\simeq b$, $\TSG$ appears to merge into $\Tc$ with showing multicritical behavior.
For larger $\Delta$, $\TSG~(=\Tc)$ becomes nearly independent of $\Delta$, being in sharp contrast to the previously-reported SG behavior, i.e., $\TSG \propto \Delta$~\cite{Saunders07, Andreanov10, Tam10}.
These peculiar SG behaviors are ascribed to reduced thermal fluctuations in the semi-discrete degenerate manifold emergent below $\Tc$.
The plateau behavior of $\TSG$ at a largely enhanced value gives an explanation for the puzzling behaviors in the pyrochlore-based antiferromagnets such as (La$_x$Y$_{1-x})$Mo$_2$O$_7$ and (Zn$_{1-x}$Cd$_x$)Cd$_2$O$_4$.

\section{Summary}
We have reviewed an extension of the loop algorithm to Heisenberg spin systems.
In \S~\ref{sec:elm}, we have explained detailed procedure of the extended loop algorithm and demonstrated its efficiency for Heisenberg spin models with single-ion anisotropy and those with biquadratic interactions.
Finally, we have reported results of our recent numerical study on spin-glass transition in a bond-disordered antiferromagnet coupled with local lattice distortions.

This report is based on the recent works done with Y. Tomita and Y. Motome.
%\bibliography{ref.bib}

\end{document}